\documentclass[prx,aps,twocolumn,reprint,floatfix,superscriptaddress,longbibliography,amssymb]{revtex4-1}
\usepackage{graphicx}
\usepackage{amsmath,amssymb}
\usepackage{accents}
\usepackage{float}
\usepackage{textcomp}
\usepackage{gensymb}
\usepackage{siunitx}
\DeclareSIUnit\gauss{G}
\usepackage[colorinlistoftodos]{todonotes}
\DeclareMathOperator*{\argmin}{argmin} 
\usepackage{xcolor}
\newcommand\hl[1]{\textcolor{black}{#1}}

\newcommand{\be}{\begin{equation}}
\newcommand{\ee}{\end{equation}}
\newcommand{\bea}{\begin{eqnarray}}
\newcommand{\eea}{\end{eqnarray}}

\newcommand{\jxone}{j^{(1)}_{x}}
\newcommand{\jyone}{j^{(1)}_{y}}
\newcommand{\jxtwo}{j^{(2)}_{x}}
\newcommand{\jytwo}{j^{(2)}_{y}}
\newcommand{\jxpone}{j^{(1)}_{x'}}
\newcommand{\jxptwo}{j^{(2)}_{x'}}

\newcommand{\rhoa}{\rho_a}
\newcommand{\rhob}{\rho_b}
\newcommand{\drho}{1-\rho_{a}/\rho_{b}}
\newcommand{\jmeas}{\delta j_y}
\newcommand{\jbulk}{j_{\mathrm{bulk}}}

\def \BFA{BaFe$_{2}$As$_{2}$}
\def \BFCA{Ba(Fe$_{1-\mathrm{x}}$Co$_{\mathrm{x}})_{2}$As$_{2}$}
\def \BFPA{BaFe$_{2}$As$_{2-x}$P$_{x}$}
\def \BFPA{BaFe$_{2}$$($As$_{1-\mathrm{x}}$P$_{\mathrm{x}}$$)$$_{2}$}

\begin{document}

\title{Imaging Nematic Transitions in Iron-Pnictide Superconductors with a Quantum Gas}

\author{Fan Yang}
\altaffiliation[F.Y,  S.F.T., and S.D.E~contributed equally to this work.]{}
\affiliation{Department of Applied Physics, Stanford University, Stanford, CA 94305, USA}
\affiliation{E.~L.~Ginzton Laboratory, Stanford University, Stanford, CA 94305, USA}
\author{Stephen F. Taylor}
\altaffiliation[F.Y,  S.F.T., and S.D.E~contributed equally to this work.]{}
\affiliation{Department of Applied Physics, Stanford University, Stanford, CA 94305, USA}
\affiliation{E.~L.~Ginzton Laboratory, Stanford University, Stanford, CA 94305, USA}
\author{Stephen D. Edkins}
\altaffiliation[F.Y,  S.F.T., and S.D.E~contributed equally to this work.]{}
\affiliation{Department of Applied Physics, Stanford University, Stanford, CA 94305, USA}
\affiliation{E.~L.~Ginzton Laboratory, Stanford University, Stanford, CA 94305, USA}
\author{Johanna C.~Palmstrom}
\affiliation{Department of Applied Physics, Stanford University, Stanford, CA 94305, USA}
\affiliation{Geballe Laboratory for Advanced Materials, Stanford University, Stanford, CA 94305, USA}
\affiliation{Stanford Institute for Materials and Energy Science, SLAC National Accelerator Laboratory,
2575 Sand Hill Road, Menlo Park, California 94025, USA}
\author{\\Ian R.~Fisher}
\affiliation{Department of Applied Physics, Stanford University, Stanford, CA 94305, USA}
\affiliation{Geballe Laboratory for Advanced Materials, Stanford University, Stanford, CA 94305, USA}
\affiliation{Stanford Institute for Materials and Energy Science, SLAC National Accelerator Laboratory,
2575 Sand Hill Road, Menlo Park, California 94025, USA}
\author{Benjamin L.~Lev}
\affiliation{Department of Applied Physics, Stanford University, Stanford, CA 94305, USA}
\affiliation{E.~L.~Ginzton Laboratory, Stanford University, Stanford, CA 94305, USA}
\affiliation{Department of Physics, Stanford University, Stanford, CA 94305, USA}

\date{\today}

\begin{abstract}
The SQCRAMscope is a recently realized Scanning Quantum CRyogenic Atom Microscope that utilizes an atomic Bose-Einstein condensate to measure  magnetic fields emanating from solid-state samples. The quantum sensor does so with unprecedented DC sensitivity at micron resolution from room-to-cryogenic temperatures~\cite{Yang:2017br}. An additional advantage of the SQCRAMscope is the preservation of optical access to the sample: Magnetometry imaging of, e.g., electron transport may be performed in concert with other imaging techniques. This \textit{multimodal} imaging capability can be brought to bear with great effect in the study of nematicity in iron-pnictide high-temperature superconductors, where the relationship between electronic and structural symmetry-breaking resulting in a nematic phase is under debate. Here, we combine the SQCRAMscope with an in situ microscope that measures optical birefringence near the surface. This enables simultaneous and spatially resolved detection of both bulk and near-surface manifestations of nematicity via transport and structural deformation channels, respectively. By performing the first \textit{local} measurement of emergent resistivity anisotropy in iron pnictides, \hl{we observe sharp, nearly concurrent  transport and structural transitions.}  More broadly, these measurements demonstrate the SQCRAMscope's ability to reveal important insights into the physics of complex quantum materials. 

\end{abstract}

\maketitle

\begin{figure*}[t!]
\includegraphics[width=\textwidth]{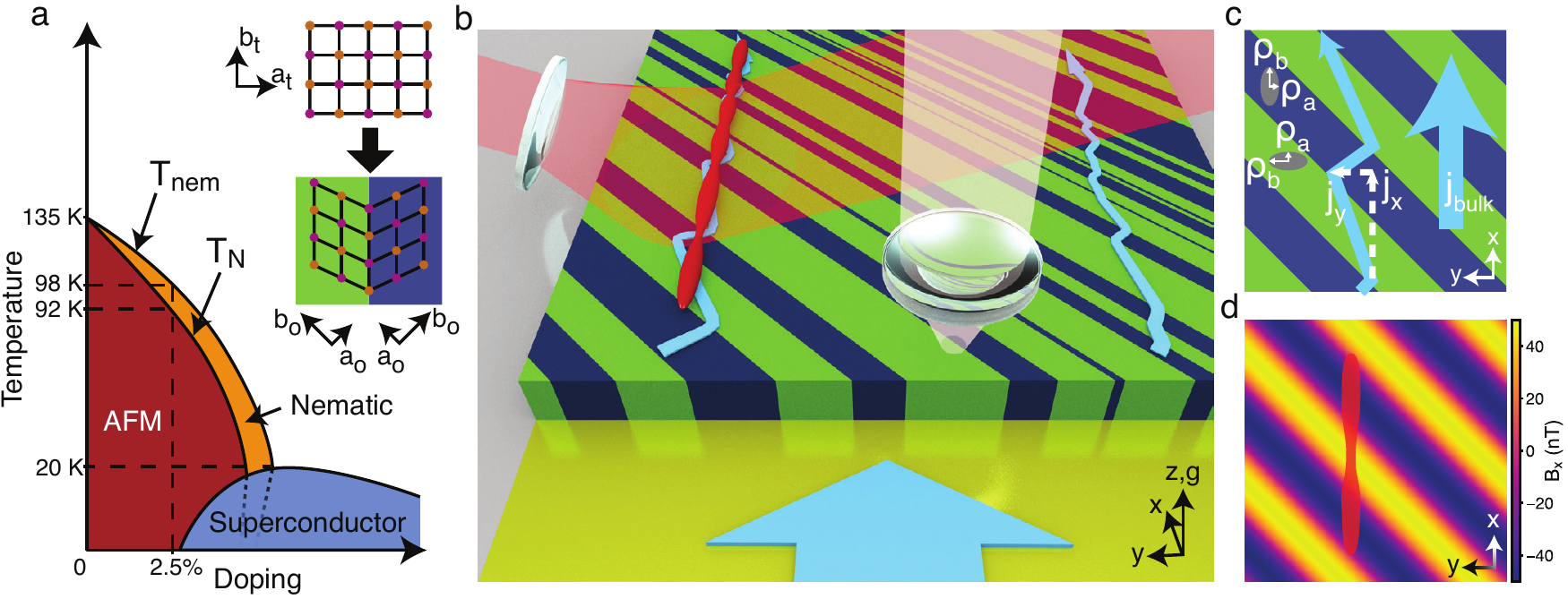}
\caption{\textbf{Multimodal SQCRAMscope.}        (a) Schematic phase diagram of \BFCA. In the high-temperature phase (white) the material is a metal with tetragonal crystal structure. At the phase boundary $T_{\text{nem}}$, the material undergoes a transition to a nematic state (orange) that breaks four-fold rotational symmetry while preserving lattice translational symmetry. At the lower temperature phase boundary $T_N$, the material becomes a stripe-like antiferromagnet (AFM, crimson). There is a dome of superconductivity (blue) that intersects the nematic transition line near its maximum critical temperature.   Inset: schematic of the crystal structure of \BFA~both above and below the structural transition.  Shown is the  Fe (brown)-As (pink) plane with the tetragonal ($t$) and orthorhombic ($o$) crystal axes labeled. (b) A quasi-1D BEC (red) is magnetically confined two-microns from the surface of the pnictide sample using an atom-chip trap (not shown)~\cite{Yang:2017br}; see Sec.~M in Supplementary Information.      The crystal forms domains with anisotropic resistivity upon cooling (blue and green stripes). Consequently, a homogeneous injection of electric current (light blue arrows) into these domains flows in a zigzag fashion from one gold contact (at bottom of panel) to the other (not shown). The density of the BEC is imaged with a high numerical aperture lens (left) by absorption imaging of a resonant laser  (transparent red).  The density modulation is proportional to the local magnetic field along $\hat{x}$ generated by the  inhomogeneous current in the sample. Vertical optical imaging through a second lens (right) using polarized light (transparent white) measures the near-surface birefringence of the crystal. The penetration depth of the imaging light is on the order of \SI{30}{\nm}, much shorter than the ${\sim}20$-$\mu$m thick  sample.  (c) In the orthorhombic phase, domains form with one of two different crystal axis orientations, shown as blue and green. These domains have an anisotropic resistivity, $\rhoa < \rhob$, as indicated by the gray ellipses. Upon crossing a domain wall, the  principal axes of the resistivity tensor are expected to interchange (as we indeed observe below). As a result, an average current density $J_{\textrm{bulk}}$ (large cyan arrow) sent through the crystal in  $\hat{x}$ bends toward $\pm\hat{y}$ at each domain boundary, forming a zigzag pattern (solid cyan path of smaller arrow).        (d) Simulation of the $x$-component of the magnetic field produced by current shown  in (c). The BEC density (red) changes according to the sign and magnitude of the $y$-component of the current density $j_y$. }
\label{fig1:cartoon}
\end{figure*}

Electronic nematicity, the breaking of crystal rotational symmetry that is driven by electronic degrees of freedom, has been intensely studied due to its proximity to high-temperature superconductivity and putative quantum criticality~\cite{Kivelson1998a,Fradkin2009,Lilly1999,Borzi2007,Daou2010,Lawler2010}. Iron-pnictide high-temperature superconductors are archetypal examples of such behavior~\cite{Fernandes2014}. \BFCA, whose schematic phase diagram is shown in Fig.~\ref{fig1:cartoon}(a), exhibits a nematic phase transition at a critical temperature $T_\text{nem}(\mathrm{x})$. Bulk structural and thermodynamic studies show that at this temperature the lattice undergoes a tetragonal-to-orthorhombic transition that spontaneously breaks four-fold ($C_4$) rotational symmetry~\cite{Rotter2008,Huang2008,Kim2011}. Simultaneous with this structural symmetry breaking is the onset of a difference in resistivity along the crystal axes, thus breaking the same $C_4$ symmetry~\cite{Chu:2010cr, Tanatar2010}. The observation of a large resistivity anisotropy in the orthorhombic state and the temperature dependence of the strain-induced resistivity anisotropy in the tetragonal state~\cite{Luo2015,Kuo958} have been interpreted as compelling evidence that the structural phase transition is driven by electronic nematic order~\cite{Fernandes:2012ja}. 

The $C_4$ rotational symmetry is broken in one of two energetically degenerate ways, resulting in twin domains~\cite{Tanatar:2009ks}; see Fig.~\ref{fig1:cartoon}(a). In the absence of a bias strain, domains of both types form with a  resistivity anisotropy that alternates in sign between neighboring domains.  Any probe that averages over a sample volume larger than the characteristic domain size will not accurately measure the behavior of a single domain. Consequently, previous bulk measurements of resistivity anisotropy have required the application of a large uniaxial  stress  to  detwin  the crystal, limiting the inferences that can be drawn about the strain-free  material~\cite{Chu:2010cr,Tanatar2010}. We overcome this by imaging bulk transport locally within domains of a nominally unstrained sample using our new SQCRAMscope technique.  Simultaneously, we directly image the domains near the surface through optical birefringence: The structural symmetry-breaking causes a rotation of the polarization of linearly polarized light upon reflection, $\theta$. This rotation also alternates in sign between domains.

Optical pump-probe spectroscopy~\cite{Thewalt2018,Stojchevska2012}, ARPES~\cite{Shimojima2014a,Sonobe2018,Yi2011}, NMR~\cite{Iye2015}, and torque magnetometry~\cite{Kasahara2012} have reported
the onset of $C_4$-symmetry breaking $\sim$20-K above $T_\text{nem}$ in \BFPA~and \BFCA, while high-resolution specific heat measurements exclude a bulk phase transition in this temperature range~\cite{Chu:2009ba,Luo2015}. The simplest explanation for this apparent contradiction is that unintended strain, e.g., from growth or mounting, induces anisotropy above $T_\text{nem}$ as a consequence of the large nematic susceptibility~\cite{Chu710}. However,  a combined micro-Laue diffraction and optical pumping study of \BFPA~indicates that the nematicity is weakest at the regions of strongest strain, casting doubt on this hypothesis~\cite{Thewalt2018}. Song~\textit{et al.}~have proposed an ``extraordinary'' nematic surface phase transition~\cite{Binder1990a,Brown2005,Song2016}, in which the surface breaks $C_4$ symmetry at a higher temperature than the bulk. Such a scenario would be consistent with the detection of anisotropy above $T_\text{nem}$ by surface sensitive probes such as ARPES and optical spectroscopy, but not in the  bulk specific heat~\cite{Luo2015}. Complicating matters, however, is that NMR~\cite{Iye2015} and torque magnetometry~\cite{Kasahara2012} detect anisotropy at $T>T_\text{nem}$ and both probes are  sensitive to the bulk. Addressing this discrepancy requires measuring the same sample simultaneously with probes that are separately sensitive to surface and bulk manifestations of nematicity.

We have recently introduced the SQCRAMscope~\cite{Yang:2017br}, a quantum-noise-limited scanning-probe magnetometer that leverages the techniques of ultracold atomic physics to image magnetism and electronic transport in solid-state samples.  The SQCRAMscope has heretofore only been tested with gold samples~\cite{Yang:2017br}; we use it for imaging strongly correlated materials for the first time in this work. By employing a magnetically levitated atomic Bose-Einstein condensate (BEC) that can be scanned within microns of the surface of a material, the microscope makes 2D spatially resolved measurements of the magnetic field emanating from samples with unprecedented DC magnetic field sensitivity.  Electronic currents flowing in the material, e.g., injected through contacts, create a magnetic field that can be imaged by the SQCRAMscope. The 2D current density can be reconstructed using the Biot-Savart law by measuring the separation between BEC and sample~\cite{Yang:2017br}. Its ability to image samples with micron resolution over a wide temperature range recommends it for the study of nematicity in pnictides.  

\begin{figure*}
    \includegraphics[width=\textwidth]{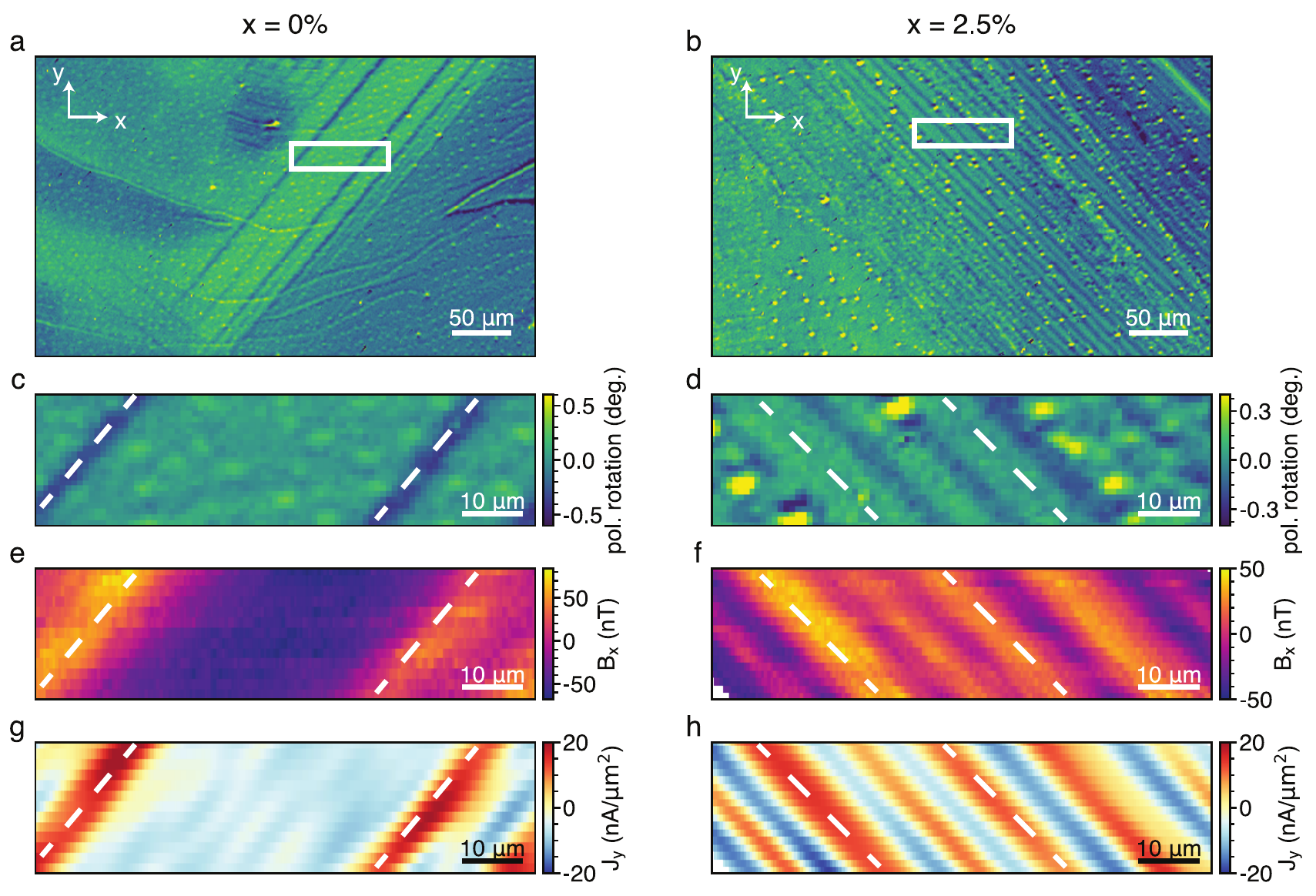}
    \caption{\textbf{Optical birefringence, magnetometry, and transport images.} Typical optical birefringence images were taken of (a) the parent compound \BFA~at \SI{92.5}{\K}, and (b) 2.5\% cobalt-doped \BFCA~at \SI{81}{\K}. The axes $x$ and $y$ are aligned with the orthorhombic crystal axes. Alternating stripes in birefringence mark the location of nematic domains. A large spatial variation of domain visibility, size, and orientation is observed in both samples. The spatial distribution of domains does not change appreciably with thermal cycling, suggesting that they are weakly pinned by  local strain introduced during the crystal growth and/or sample preparation/mounting. Typical SQCRAMscope scans were taken in regions marked by white boxes in (a) and (b), for which the optical birefringence maps are shown in (c) and (d). The magnetometry scans are shown in (e) and (f), and the reconstructed current densities are shown in (g) and (h); see Supp.~Sec.~F for reconstruction procedure. 
    The dashed lines in panels (c) and (d) are drawn at the center of two negative $\theta$ domains in (c) and two positive domains in (d).  The  lines are overlaid at the same positions in the  magnetometry and transport images as guides to the eye. }
    \label{fig2}
\end{figure*}

\begin{figure*}[t!]
    \includegraphics[width=0.99\textwidth]{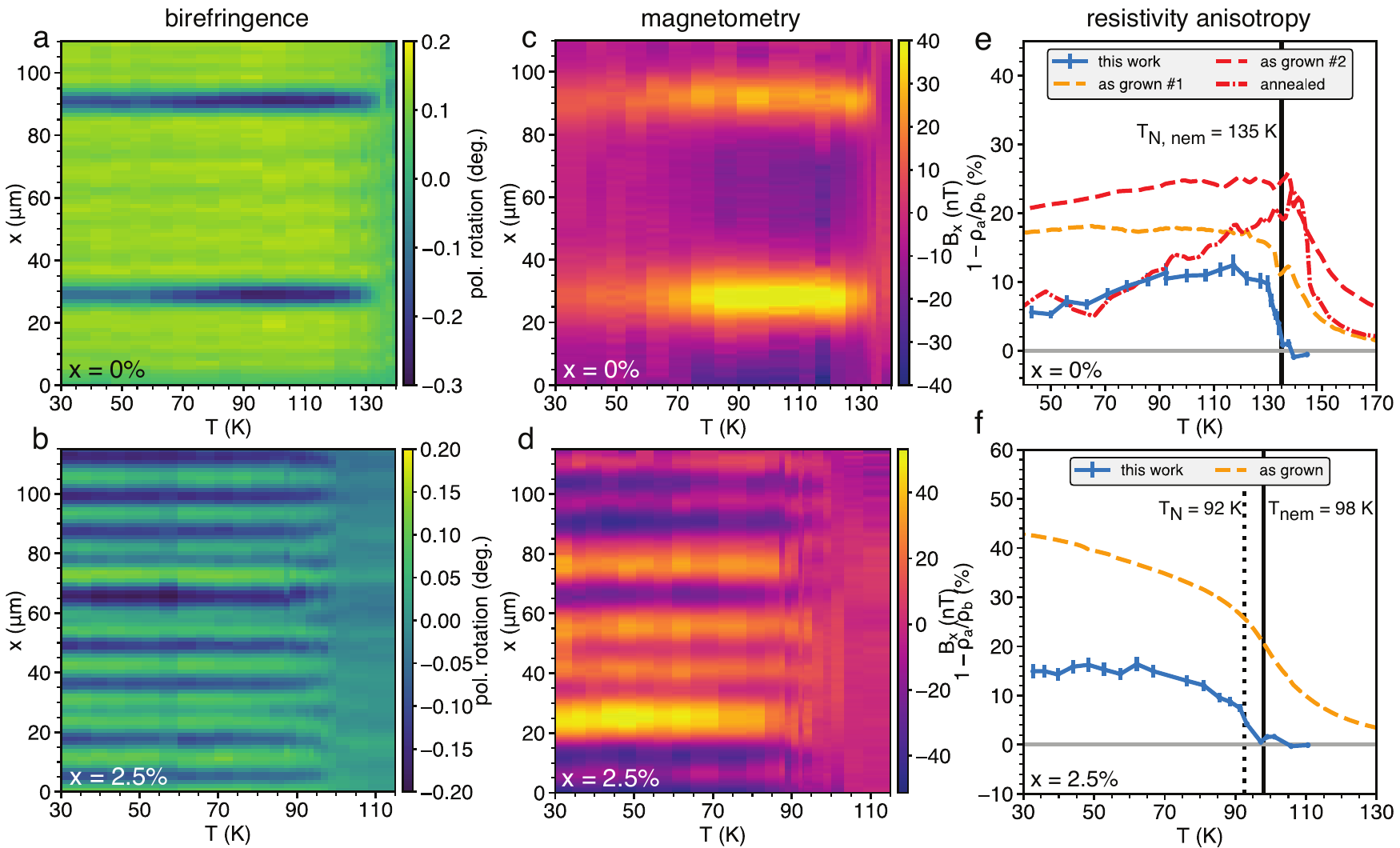}
    \caption{\textbf{Temperature dependence of nematic order.}
        (a) and (b), Birefringence $\bar{\theta}(x,T)$ for region marked P1 (D1) on $\mathrm{x}=0$\% ($\mathrm{x}=2.5$\%) sample in Supp.~Fig.~3.
        (c) and (d), Magnetic field $\bar{B_{x}}(x,T)$ for same region marked P1 (D1).
        (e) and (f), Temperature dependence of resistivity anisotropy $\drho$ calculated from SQCRAMscope data  (blue) in (c) and (d), respectively. Error bars represent both random and systematic uncertainty in resistivity anisotropy; see Supp.~Sec.~H. For comparison, we also plot bulk resistivity anisotropy under uniaxial stress for \BFCA~samples. Data for the orange curve is reproduced from~\cite{Chu:2010cr}, while  the red curves are reproduced from~\cite{ishida2013anisotropy}. The dashed (dash-dotted) curves are for ``as-grown''  (annealed) samples.   Nematic and N\'{e}el transition temperatures are indicated by solid and dotted black lines. Error bars in (e,f) represent standard error of the mean.
    }
    \label{fig3}
\end{figure*}

Figure~\ref{fig1:cartoon}(b) shows the operating principle of the SQCRAMscope. A sample, attached to a thin silicon substrate and electrically contacted by gold wires on two sides, is brought in close proximity to an atom chip. The atom chip provides a smooth, harmonic trapping magnetic field that confines a quasi-1D BEC (red) within microns of the surface of the sample. \hl{See Supp.~Sec.~M for trap characterization.} Any magnetic field $\mathbf{B}(x,y)$ sourced from the sample is superposed upon that of the trap. Along the long, weakly confined axis of the BEC ($\hat{x}$), the density of the BEC will respond to the $x$-component of the magnetic field $B_x(x,y)$. By imaging the density of the BEC with a high-numerical-aperture (NA) lens while scanning the sample position with respect to the BEC, we can create 2D maps of $B_x(x,y)$. In this experiment, $B_x$ is produced by spatial inhomogeneity of one component of the current density $j_y(x,y)$ flowing through the sample. Deconvolving $B_x(x,y)$ with the Biot-Savart kernel allows one to calculate $j_y(x,y)$ from the measured field $B_x(x,y)$; see Supp.~Sec.~F. The SQCRAMscope's current density resolution is bounded from below by the resolution in magnetic field. This in turn is limited by the NA of the imaging objective to 2.2~$\mu$m~\cite{Yang:2017br}. The effective resolution is larger for samples thicker than a few microns; see Supp.~Sec.~F.

We augmented the SQCRAMscope with an in situ optical birefringence microscope; see rightward lens in Fig.~\ref{fig1:cartoon}(b) and description in Supp.~Sec.~K. In the context of the iron pnictides, the angle $\theta(x,y)$ by which the polarization of linearly polarized light is rotated upon reflection is, to first order, linearly proportional to the orthorhombic structural distortion.  Birefringence measurements have previously been used to image twin domain formation~\cite{Tanatar:2009ks,Prozorov:2009ba}. The \SI[number-unit-product=-]{780}{\nm} light used in our birefringence microscope has a skin depth of \SI{30}{\nm}~\cite{Stojchevska2012} in \BFCA~and is thus primarily sensitive to structure near the surface of the material.  In contrast, SQCRAMscope magnetometry is sensitive to the current density convolved throughout the bulk of the sample; see Supp.~Sec.~F. This \textit{multimodal} SQCRAMscope thus has selective sensitivity to both bulk and near-surface nematicity.

We now present magnetometry and birefringence measurements of single-crystal \BFCA~with dopings $\mathrm{x}=0$ ($T_\text{nem} = \SI{135}{\K}$) and 2.5\% ($T_\text{nem} = \SI{98}{\K}$); see Supp.~Sec.~C. These crystals are known to form domain walls along the $\langle 100 \rangle$ directions of the tetragonal Fe lattice. For the samples we study, cleaved ${\sim}25$-$\mu$m thick and cut along the $\langle 110 \rangle$  directions, we expect domains to form at the temperature of rotational symmetry breaking with domain walls oriented at \ang{45} with respect to the sample edges---e.g., \ang{45} to $\hat{x}$ and $\hat{y}$ in Figs.~\ref{fig1:cartoon}(b--d)---and no bulk $\hat{z}$ dependence~\cite{Tanatar:2009ks}.

The domains will manifest as alternating stripes in birefringence measurements where $\theta(x,y)$ changes sign, as shown schematically in Fig.~\ref{fig1:cartoon}(c). The domain structure will also present itself in the meandering of current density flowing through the domains due to the abrupt change in the resistivity tensor at the domain walls. On one side, $\rho_{x} > \rho_{y} $, and vice versa on the other. Current, injected by macroscopic contacts and oriented along  $\hat{x}$, will take the path of least resistance through this domain structure. As shown in Fig.~\ref{fig1:cartoon}(c), $j_y(x,y)$ will change sign at each domain wall while $\mathbf{j}(x,y)$, spatially averaged over all domains, yields a net current $j_\mathrm{bulk}$ along $\hat{x}$. $j_y(x,y)$ creates a spatially modulated $B_x(x,y)$ that follows the underlying domain structure. Figure~\ref{fig1:cartoon}(d) shows the simulated $B_x(x,y)$ that would arise from the domain structure in Fig.~\ref{fig1:cartoon}(c); see Supp.~Sec.~J. Thus, in a SQCRAMscope measurement wherein the BEC is oriented along $\hat{x}$, the expected signatures of nematic domains are peaks and valleys in $B_x(x,y)$ in correspondence with those in $\theta(x,y)$.

\begin{figure*}
    \includegraphics[width=0.995\textwidth]{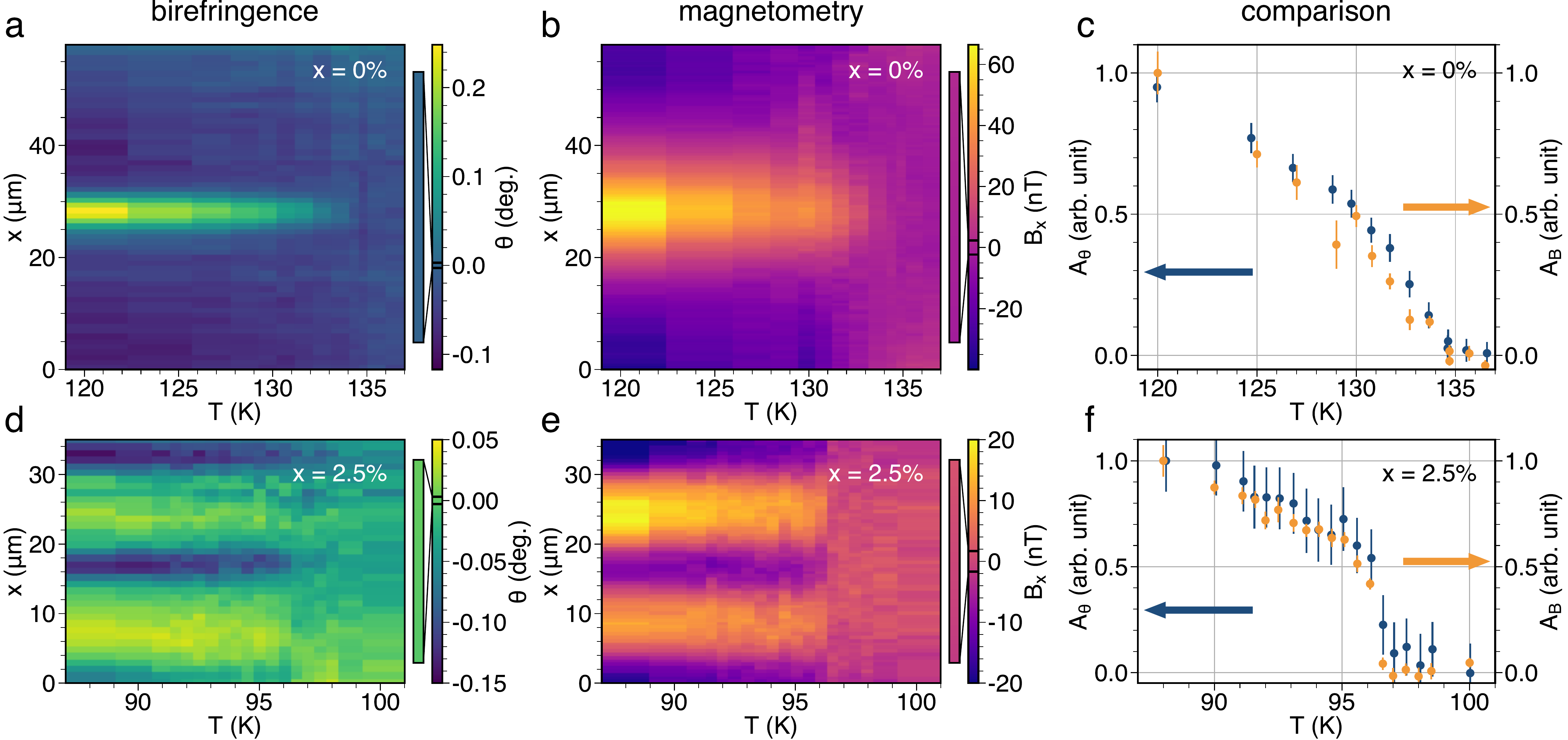}
    \caption{\textbf{Nematic transition temperatures.} We directly compare the temperature dependence of the polarimetry and the magnetometry data. One representative dataset is shown for  an $\mathrm{x}=0$ sample at location P2 in panels (a--c), and another for an $\mathrm{x}=2.5$\% sample at location D2 in panels (d--f); see Supp.~Sec.~D for location pictures. Shown in columns from left to right are (a,d) the birefringence signal during sample warm-up through the transition, (b,e) the magnetometry signal taken concurrently from the same region, and (c,f) the domain amplitudes $A_{B}(T)$ (orange) and $A_{\theta}(T)$ (blue).   \hl{The expanded color bars represent the measurement error around zero for (a,d) birefringence and (b,e) magnetometry. The nematic domains disappear in polarimetry and the magnetometry at roughly the same temperature for both the parent and doped compounds. Error bars in (e,f) represent standard error of the mean.} }
    \label{fig4}
\end{figure*}

Figures~\ref{fig2}(a) and \ref{fig2}(b) show large area $\theta(x,y)$ images for $x = 0$ and $\mathrm{x}=2.5$\% at $T=\SI{92.5}{\K}$ and $T=\SI{81}{\K}$, respectively. Both clearly show the formation of twin domains oriented along the $\langle 110 \rangle$ directions. The remaining panels in Figure~\ref{fig2} show magnified $\theta(x,y)$, $B_x(x,y)$, and $j_y(x,y)$ images corresponding to the areas marked in white in panels (a) and (b). As expected, both $B_x(x,y)$ and $j_y(x,y)$ exhibit stripes in correspondence with those in $\theta(x,y)$.

We measured $B_x(x,y)$ and $\theta(x,y)$ for temperatures between \SIrange{36}{138}{\K} for $\mathrm{x}=0$ and  \SIrange{33}{110}{\K} for $\mathrm{x}=2.5$\%.  Images are taken in the  same fields of view as in Figs.~\ref{fig2}(c,d); see scan regions P1 and D1 in Supp.~Sec.~D.   We average $B_x(x,y)$ and $\theta(x,y)$ along the translationally invariant direction of the twin domain walls at each temperature. The resulting functions $\bar{B}_x(x,T)$ and $\bar{\theta}(x,T)$ are plotted in Figs.~\ref{fig3}(a--d).
At each temperature, the resistivity anisotropy is calculated from the current density using 
$\drho = 2 \jmeas/(\jbulk + \jmeas)$, 
where $\jmeas$ is the change in $j_y$ between adjacent domains and  is calculated from $\bar{B}_x(x,T)$; see Supp.~Sec.~H. The total current density $j_\mathrm{bulk}$ is calculated from the sample geometry and known sample current.  This value is consistent with its effect on the trapping potential of the BEC; see Supp.~Sec.~E. 

We plot the temperature dependence of the resistivity anisotropy, $\drho$, for $\mathrm{x}=0$\% and 2.5\% in Figs.~\ref{fig3}(e,f). These represent the first \textit{local} measurements of resistivity anisotropy under nominally strain-free conditions. For comparison, we also plot the resistivity anisotropy for samples of the same doping, but measured using bulk resistivity under uniaxial stress~\cite{Chu:2010cr,ishida2013anisotropy}. The onset of resistivity anisotropy is sharper for the local SQCRAMscope measurements and occurs closer to $T_{\text{nem}}$. This can be attributed to the large symmetry-breaking strain applied to detwin the crystals in the bulk measurements that, in the presence of a large nematic susceptibility~\cite{Chu710}, results in significant resistivity anisotropy above $T_{\text{nem}}$. The sharpness of the transition in the local SQCRAMscope measurement empirically demonstrates the nearly strain-free conditions in our experiment.

We observe a local resistivity anisotropy that is generally smaller than that from these previously reported bulk measurements. This discrepancy deserves comment. First, a difference between our nominally strain-free measurements and those under uniaxial strain is expected near $T_\text{nem}$ due to the diverging nematic susceptibility~\cite{Chu710}. That this discrepancy persists to the lower temperatures is more surprising.  Figure~\ref{fig3}(e) shows that our measured resistivity anisotropy at low temperatures is, to within error, the same as for one of these strained samples. While comparable data is not available for $\mathrm{x}=2.5$\%, this suggests that the discrepancy between our measurements and those of strained samples is within the range of sample-to-sample variation. Another possibility is that the apparent resistivity anisotropy in the SQCRAMscope measurement is smaller due to the presence of domains smaller than our spatial resolution~\cite{Ma2009}. However, it may also be that the resistivity anisotropy in unstrained samples is, in fact, systematically smaller than in those under uniaxial strain. This merits further investigation beyond the scope of this paper.

We now turn to a comparison of the temperature dependencies of the magnetic versus optical signatures of rotational symmetry breaking. To track the onset of nematic domain formation, we define the domain-averaged amplitude for magnetometry and birefringence modulations to be $A_B(T)$ and $A_{\theta}(T)$, respectively; see~Supp.~Sec.~I. Figures~\ref{fig4}(a,b) show the temperature dependencies of $\bar{\theta}(x,T)$ and $\bar{B}_x(x,T)$ for the region P2 in the $\mathrm{x}=0$ sample. The resulting $A_{\theta}(T)$ and $A_{B}(T)$ are plotted in Fig.~\ref{fig4}(c). The signal in both channels drops to zero at \SI{135}{\K}, in agreement with the measured bulk nematic transition temperature $T_\text{nem}$~\cite{Chu:2010cr}.

Figures~\ref{fig4}(d--f) show analogous data for region D2 in the $\mathrm{x}=2.5$\% sample.    As shown in panel (f),  $A_{\theta}(T)$ and $A_{B}(T)$ exhibit a sharp change around 96.5~K\hl{, a temperature consistent with transport measurements~\cite{Chu:2010cr}.  While nearing the limit of our sensitivity, $A_{\theta}(T)$  maintains a very small positive value up to $\sim$100~K, which may be due to intrinsic strain coupled to nematic susceptibility.  The nearly concurrent transitions point to the absence of an extraordinary surface transition at the short length scales defined by these domain sizes. We cannot yet comment upon whether an extraordinary surface transition is the cause of the \textit{longer-length-scale} $C_4$ symmetry-breaking, typically observed up to 20-K above $T_\text{nem}$~\cite{Thewalt2018,Stojchevska2012,Shimojima2014a,Sonobe2018,Yi2011,Iye2015,Kasahara2012}. However, future improvements to the SQCRAMscope field-of-view would allow us to observe whether long-length-scale bulk anisotropic resistivity  coincides with a symmetry-broken state manifested in birefringence features. In an enabling step toward such future measurements, we have confirmed that our samples also exhibit long-length-scale birefringence, indeed extending as high as 60-K above $T_\text{nem}$; see Supp.~Sec.~A. Last, we note that domain wall movement near the transition is observed in other regions of the $\mathrm{x}=2.5$\% sample; see Supp.~Sec.~B for data and analysis. This also warrants future investigation.}

In summary, we have made the first local resistivity anisotropy measurements in iron pnictides by using SQCRAMscope magnetometry. \hl{We find that nominally strain-free \BFCA~crystals exhibit a sharper nematic transition than that of uniaxially biased bulk crystals.  Moreover, we perform simultaneous measurements of optical rotation arising from structural distortion near the surface. Comparison of the two measurements reveal no sign of $C_4$-symmetry breaking above the onset of resistivity anisotropy at $T_\textrm{nem}$ for the short-length-scale domains observed.} The local measurements presented here demonstrate the versatility of the SQCRAMscope as a novel quantum sensor for studying quantum materials. By combining ultrasensitive magnetometry with room-to-cryogenic temperature operation, micron resolution, and optical access for complementary imaging modalities, the SQCRAMscope is now well placed to image a wide range of quantum materials, including high-$T_c$ cuprates and electron hydrodynamic materials.

We thank S.~Kivelson for enlightening discussions, and J.-H. Chu for early samples. We acknowledge funding support for apparatus construction from ONR (N00014-17-1-2248). Funding for F.Y.~and partial support for S.D.E.~was provided by the U.S.~Department of Energy, Office of Science, Office of Basic Energy Sciences, under Award Number DE-SC0019174. Crystal growth and sample preparation were supported by the Department of Energy, Office of Basic Energy Sciences, under Contract No.~DE-AC02-76SF00515.   Fabrication of sample mount substrates and atom chip were performed at the Stanford Nanofabrication Facility and the Stanford Nano Shared Facility, supported by the NSF under award ECCS-1542152.  S.D.E.~acknowledges partial support from the Karel Urbanek Postdoctoral Fellowship.  S.F.T. and B.L.L.~acknowledge support from the Gordon and Betty Moore Foundation through Grant No.~GBMF3502 and from the ARO (W911NF1910392).  J.C.P.~acknowledges support from an NSF Graduate Research Fellowship (DGE-114747), a Gabilan Stanford Graduate Fellowship, and the Gerald J.~Lieberman Fellowship.

\clearpage
\pagebreak

\section*{Supplemental Material}

\setcounter{figure}{0}  

\renewcommand{\figurename}{Supp.~Fig.}

\begin{figure*}
    \includegraphics[width=0.99\textwidth]{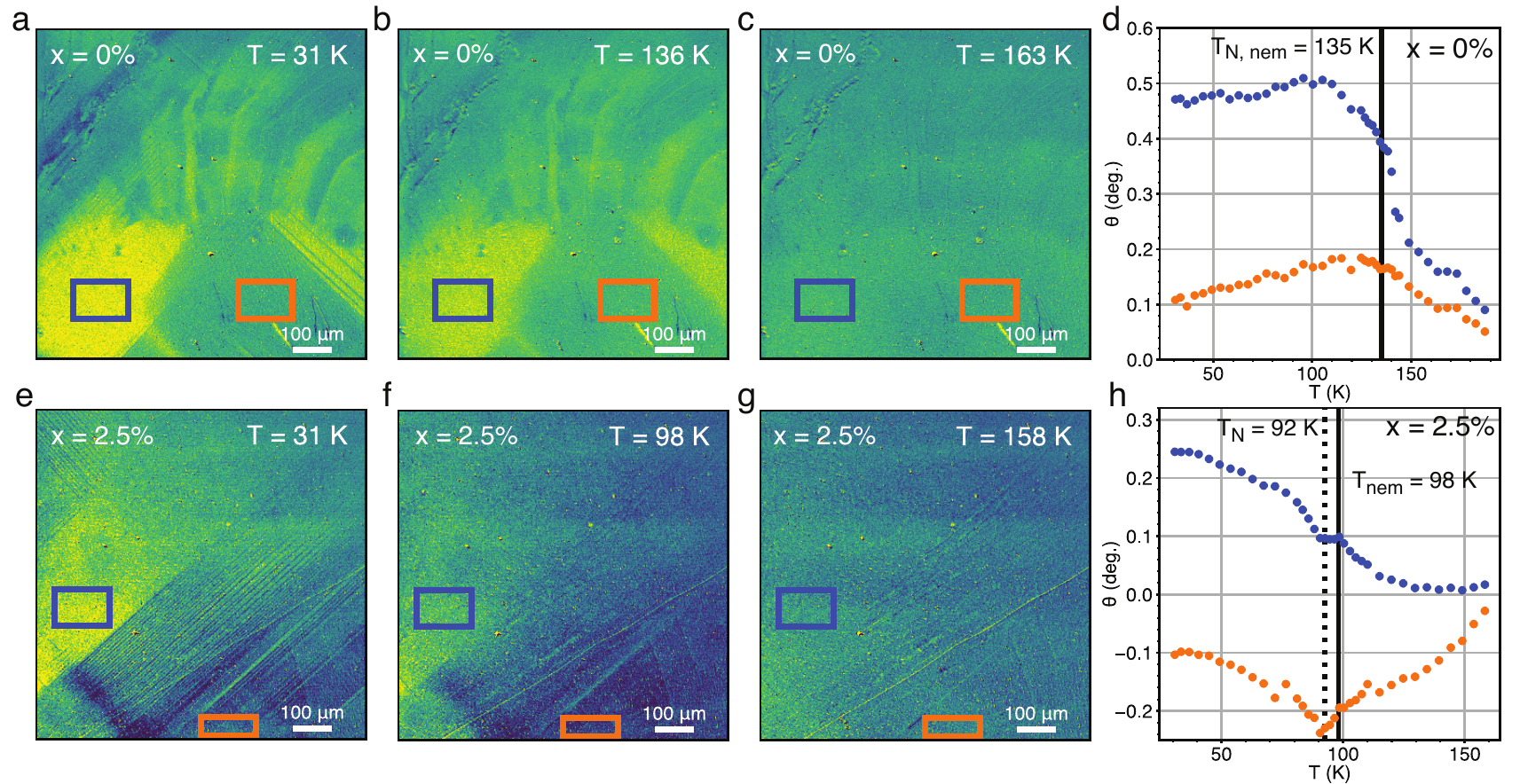}
    \caption{{Large-length-scale optical birefringence modulation. }
    Data from $\mathrm{x}=0$\% and 2.5\%-doped samples in panels (a-d) and (e-h), respectively. 
    Optical birefringence  below the nematic transition (a,e), at the transition (b,f), and above the transition (c,g). (d,h) Polarization rotation angle as a function of temperature at two different regions marked by blue and orange rectangles in (a-c) and (e-g), respectively.   Nematic and N\'{e}el transition temperatures are indicated by solid and dotted lines.
    }
    \label{fig5}
\end{figure*}

\subsection{Long-length-scale inhomogeneous birefringence}
\label{ssec:lls_biref}

A variety of experimental probes are reported as having detected rotational $C_4$ symmetry breaking persisting tens of kelvin above the bulk structural transition temperature (as determined by scattering and thermodynamic probes)~\cite{Thewalt2018,Stojchevska2012,Shimojima2014a,Sonobe2018,Yi2011,Iye2015,Kasahara2012}. \hl{However, our local probes  are consistent with the concurrent onset of resistivity anisotropy and twin-domain formation, at least where the domains walls do not wander; see Supp.~Sec.~\ref{ssec:additional_data}.  } To reconcile this apparent discrepancy, we now detail our observation of birefringence that is inhomogeneous on long-length-scales and persists to temperatures as high as \SI[number-unit-product=-]{60}{\K} above $T_{\text{nem}}$. These observations are consistent with the previous reports of $C_4$ symmetry breaking at $T > T_{\text{nem}}$.  

Supplemental Fig.~\ref{fig5} shows wide-area birefringence ($\theta(x,y)$) images for $\mathrm{x}=0$ and 2.5\% samples at temperatures below, at, and above their respective $T_\text{nem}$. While short-length-scale birefringence modulations from twin-domains disappear above  $T_\text{nem}$,  long-length-scale, inhomogeneous birefringence persists up to \SI[number-unit-product=-]{60}{\K} higher.   

Supplemental Figs.~\ref{fig5}(d,h) show  birefringence in the blue and orange boxes  plotted versus temperature. Evident are the expected kinks in $|\theta(x,y)|$ due to the peak in nematic susceptibility near $T_\text{nem}$. Interestingly, the sign of $\theta(x,y)$ in the parent sample is the same in the two regions despite being morphologically consistent with orthogonally oriented domains. This might indicate the presence of a quenched lattice distortion from unintended strain along the same direction in both regions. That could cause the ${\sim}{+}0.2$-baseline shift we observe around which short and long-length-scale birefringence is modulated.  Such a picture is also consistent with the observed peak, rather than dip, in the orange region's rotation, since the prevailing distortion axis would be co-aligned with that in the blue region.

We were not able to determine whether long-length-scale resistivity anisotropy is coincident with this birefringence \hl{due the currently limited field of view of the SQCRAMscope}. Nevertheless, these observations suggest that the anisotropy observed by other probes~\cite{Thewalt2018,Stojchevska2012, Shimojima2014a,Sonobe2018,Yi2011, Iye2015,Kasahara2012} at $T>T_\text{nem}$ may be the result of a large nematic susceptibility coupled to inhomogeneous unintended strain.

\subsection{\hl{Additional scan of $\mathrm{x}=2.5$\% sample}}
\label{ssec:additional_data}

\begin{figure*}
    \centering
    \includegraphics[width=0.99\linewidth]{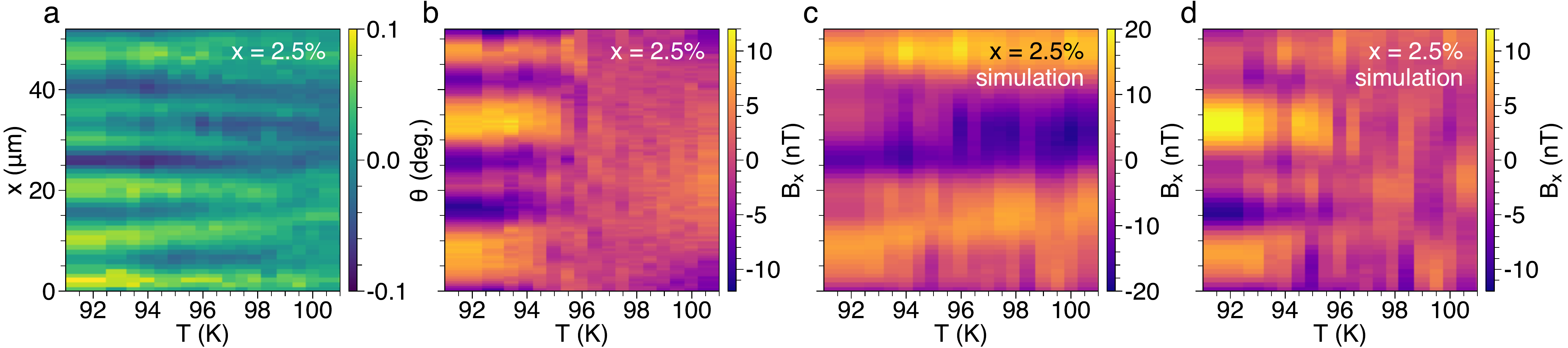}
    \caption{Additional scan of the  $\mathrm{x}=2.5$\% sample in D3 region. Shown are (a) birefringence signal $\bar{\theta}(x, T)$, (b) magnetometry signal $\bar{B_{x}}(x, T)$, and (c)  simulated magnetic field $\bar{B}^{\mathrm{sim}}_{x}(x, T)$ from the birefringence data in (a), and (d) simulated magnetic field with background subtracted $\bar{B}^{\mathrm{sim}}_{x}(x, T) - \bar{B}^{\mathrm{sim}}_{x}(x, T^*)$ ($T^* > T_\mathrm{nem})$.}
    \label{supfig:simulation}
\end{figure*}

\hl{Supplementary Fig.~\ref{supfig:simulation}(a,b) shows birefringence and measured magnetic field for scan region D3 of the  $\mathrm{x}=2.5$\% sample. The scan exhibits interesting features, highlighting the variety of domain morphology at the nematic transition.  Limited magnetometry resolution blurs neighboring domains that otherwise appear distinct in optical birefringence. Birefringence measurements reveal the movement of domain walls near the transition temperature that can just  be discerned in the lower-resolution magnetometry data, implying a bulk, rather than just near-surface effect. This movement, coupled with sensitivity and resolution limits, makes it difficult to compare the  temperatures at which  transitions in birefringence versus magnetometry occur in this scan.  Rather than attempt to compare metrics like  $A_B(T)$ and $A_{\theta}(T)$ for migrating domains of variable width, we instead simulate the magnetic field resulting from the domain structure in panel (a) and use it to compare with the actual magnetometry data in panel (b).  To do so, we calculate the field we \textit{would} detect if the birefringence structure in panel (a) were to faithfully serve as proxy for demarcating the bulk electronic nematic domain boundaries.  The results are shown in panel (d), where the optical data have been convolved with the Biot-Savart kernel to account for finite resolution and noise from the sensitivity limit measured in Ref.~\cite{Yang:2017br} has been added; see below for simulation details.  }

Simulated magnetometry domains appear to be subsumed by noise above 97~K, nearly the same temperature at which the domains in magnetometry vanish outright or are also subsumed into detection noise. Because of this, we cannot tell if the magnetometry domains actually extend to higher temperatures, say, up to $\sim$100~K where the optical domains vanish.  However, domains in the magnetometry data do seem to migrate just as in the simulation, showing that this is an effect of the bulk as well as the surface. Another complication is the broad magnetic feature of unknown origin that appears above $\sim$97~K, which could also be masking the magnetic domain signal. The positive part of the feature is centered $\pm15$-$\mu$m about $x = 20$~$\mu$m, while negative wings emerge at the top and bottom of the scan.  This feature is not evident in the optical measurement, though it is above the noise floor and presumably not a detection artifact.  

We cannot discern whether domains in magnetometry  vanish around 97~K due to detection noise or because they are subsumed into a broad magnetic feature of unknown origin.  Thus, while we cannot be sure whether the $\mathrm{x}=2.5$\% data is more consistent with a \textit{single} nematic structural-electronic transition versus an extraordinary surface transition scenario, the data do indicate that bulk and surface transitions differ by no more than few degrees and that strong short-length-scale nematic domains do not extend far above the expected bulk nematic transition temperature. Future magnetometry using a SQCRAMscope with improved resolution and sensitivity should help resolve what this feature is and exactly where the bulk transition lies in data where the domains migrate.

\subsubsection*{\hl{Simulation of expected magnetic field from optical rotation data}}

\hl{We will show in Supp.~Sec.~\ref{ssec:ra_model} that to leading order, the current density $j_{y}(x, y)$ is proportional to the resistivity anisotropy $1 - \rho_x / \rho_y$. Therefore, if we take the birefringence $\theta(x, y)$ as a proxy for nematicity, and by extension, resistivity anisotropy, then convolving it with the Biot-Savart kernel $G(k_x, k_y)$  gives the expected magnetic field $\bar{B}^{\mathrm{sim}}$; see Eq. \eqref{equ:bskernel} in Supp.~Sec.~\ref{ssec:deconv}.}

\hl{The raw and background-subtracted simulation results are shown in Supp.~Figs.~\ref{supfig:simulation}(c,d). We now detail the numerical procedure for generating the simulated field below. Supplemental~Sec.~\ref{ssec:ra_model} presents how to map measured current density to resistivity anisotropy.  Briefly,  the current density $j_{y}(x, y)$ perpendicular to the BEC is proportional, to leading order, to the resistivity anisotropy $\drho$,
\begin{align}
\drho = \frac{2 \jmeas}{\jbulk},
\label{eq:anis}
\end{align}
where $\jbulk$ is the applied bulk current density. Therefore, if we take the birefringence $\theta(x, y)$ as a proxy for nematicity, and by extension, resistivity anisotropy,
\begin{align}
    \drho = \alpha \theta(x, y),
\end{align}
where $\alpha$ is an arbitrarily chosen conversion coefficient,
then convolving $\jmeas$ with the Biot-Savart kernel $G(x, y)$  (see Supp.~Eq.~\ref{equ:bskernel} in Supp.~Sec.~\ref{ssec:deconv}) gives the expected magnetic field $B_x^{\mathrm{sim}}$:
\begin{align}
    B_x^{\mathrm{sim}}(x, y) & = j_y^\mathrm{sim}(x, y) * G(x, y) \\
    & = \frac{\alpha \jbulk}{2} \theta(x, y) * G(x, y).
\end{align}}

\hl{We note that the application of the aforementioned method to the data in Supp.~Fig.~\ref{supfig:simulation}(a) results in Supp.~Fig.~\ref{supfig:simulation}(c), which does not fully resemble the measured field shown in Supp.~Fig.~\ref{supfig:simulation}(b). However, by subtracting the simulated field at a higher temperature $T^* = \SI{102.5}{\K}$ from that of each temperature, we recover a field map Supp.~Fig.~\ref{supfig:simulation}(d) that is more similar to the measured field. This suggests 
that the long-length-scale birefringence reported in Supp.~Sec.~A might contribute an offset to the data observed here.}

\begin{figure*}[t!]
    \centering
    \includegraphics[width=0.99\linewidth]{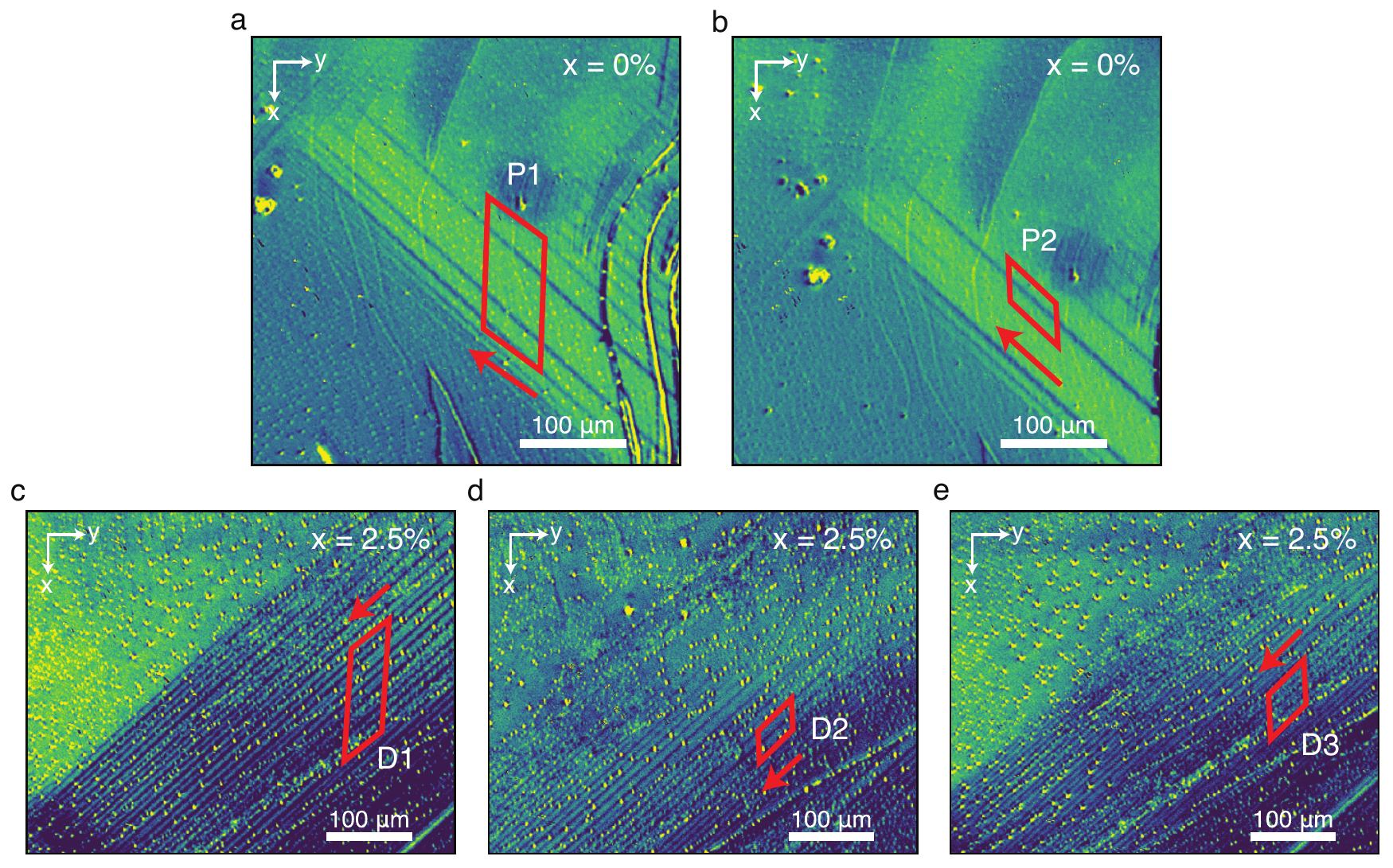}
    \caption{SQCRAMscope magnetometry scan regions. The scan regions P1 and P2 are indicated by red parallelograms on $\theta(x,y)$ images for parent \BFA~in panels (a--b). The regions D1-3 are similarly indicated for 2.5\% Co-doped \BFCA~in panels (c--e). Red arrows indicate the direction of the scan. The speckles seen on the sample surfaces were likely introduced post-growth via accidental ablation of glue. They do not have a noticeable effect on electronic transport.}
    \label{supfig:domain_loc}
\end{figure*}

\begin{figure}[t!]
    \centering
    \includegraphics[width=0.99\linewidth]{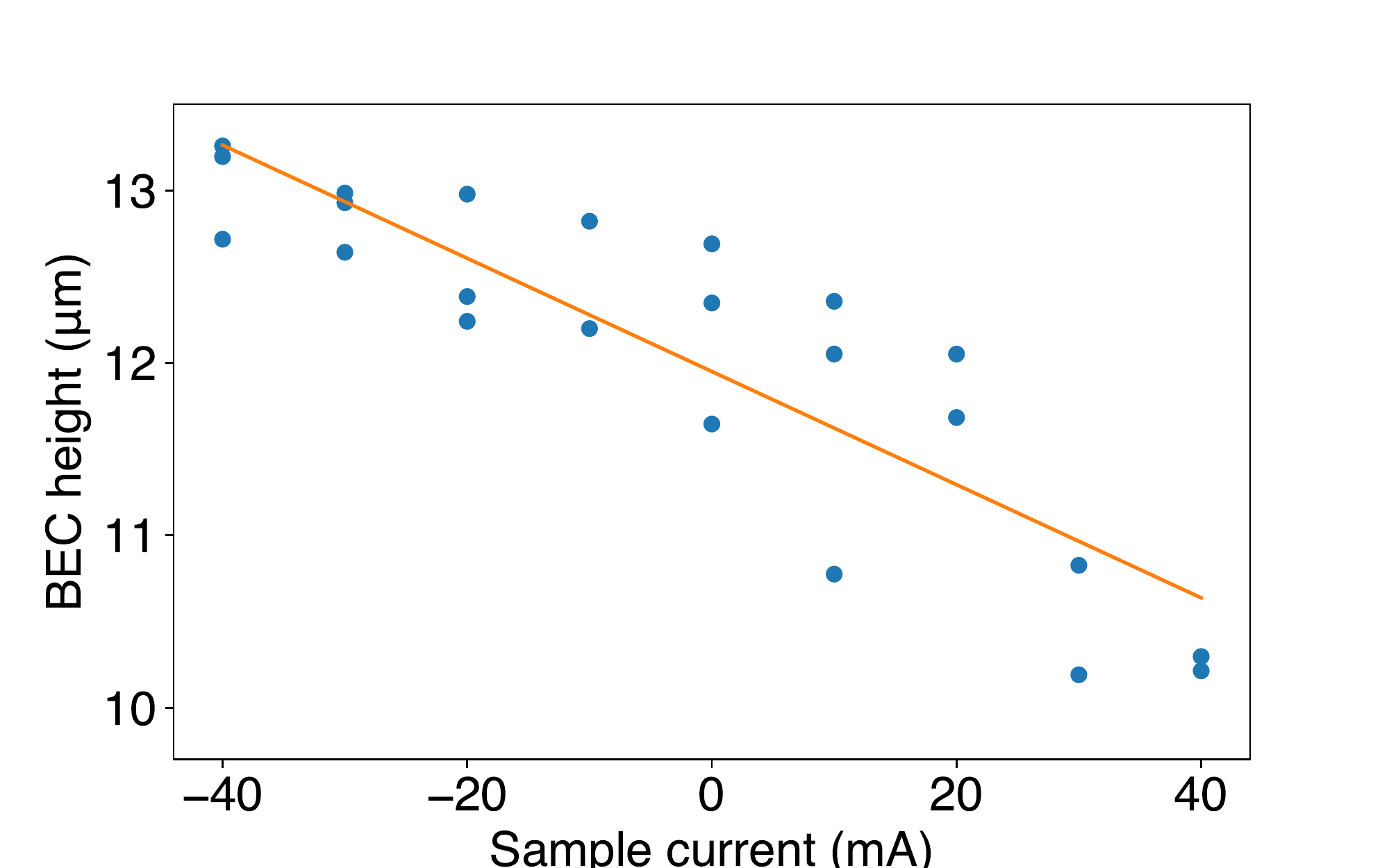}
    \caption{BEC height versus sample current.  Error bars represent standard error of the mean.}
    \label{fig:vertscan}
\end{figure}

\vspace{3mm}
\subsection{Sample preparation}
\label{ssec:sprep}
Single crystals of \BFCA, with nominal composition $\textrm{x}=0$\% and 2.5\%, were grown using the self-flux technique described in Ref.~\cite{Chu:2009ba}. The structural and N\'{e}el transition temperatures were determined from bulk resistivity measurements on crystals from the same growth batch following the procedure in Ref.~\cite{Chu:2010cr}. The intra-batch variations in transition temperatures are typically \SI{\sim 1}{\K}.

The crystals were cleaved and cut into thin rectangular plates, with edges of the crystal cut at roughly \ang{45} to the tetragonal axis. The sizes of the crystals were measured using an SEM to be:  $\SI{1.69}{\mm} \times \SI{2.35}{\mm} \times \SI{28}{\um}$ for the parent crystal, and $\SI{1.78}{\mm} \times \SI{2.23}{\mm} \times \SI{22}{\um}$ for the 2.5\%-doped crystal. The variation in thickness is of the order \SIrange{5}{10}\%. The crystals were positioned on lithographically patterned gold wires on the silicon wafer used to support the samples in the SQCRAMscope using a flip-chip bonder, and electrical contact between crystal and gold was made using silver epoxy.

\subsection{Location of scan regions}
\label{ssec:scan_regions}

The location of the scan regions P1, P2, D1, D2 and D3 referred to in the main text are indicated in Supp.~Fig.~\ref{supfig:domain_loc}.

\subsection{Magnetometry measurement of bulk current density}
\label{ssec:j_bulk}


In our calculations of resistivity anisotropy, we use a bulk current density $j_{\textrm{bulk}}$ that is determined by dividing the total sample current by the cross-sectional area of the sample. This assumes that $j_{\textrm{bulk}}$ is spatially homogeneous.  To substantiate this assumption, we carried out a more direct local measurement of $j_{\textrm{bulk}}$, as we now explain. (The data listed below are taken from the parent compound as an example. Similar measurements carried out on the 2.5\%-doped sample shows no discrepancy and are omitted here.) Due to the shape of our BEC trapping field, applying a magnetic field perpendicular to the BEC changes the distance between the sample surface and the BEC (which we refer to as the BEC height). The bulk sample current flows parallel to the BEC (along the $x$ axis), which generates a perpendicular magnetic field (along the $y$ axis). Thus, the  BEC height reflects the mean current density $\bar{j}_\textrm{bulk}$ in its vicinity. 

Starting with the BEC positioned about \SI[number-unit-product=-]{10}{\um} away from the sample surface, we measured its height as a function of the total sample current, as shown in Supp.~Fig.~\ref{fig:vertscan}. We can determine the response of the gas height to the sample current to be \SI{-37 \pm 5}{\um / \A} in this example measurement.
We then calibrated the response with external bias coils to find a coefficient of \SI{11\pm 1}{\um / \gauss}, from which we deduce that the sample generates a bias field near its surface with a field-per-current coefficient of $B_x / I = \SI{3.4\pm 0.5}{\gauss / \A}$. Since the BEC height is much smaller than the lateral size of the sample, the field near the sample can be modelled by a thick infinite-sized slab conductor with current density $j_y$ and magnetic field $B_x$ given by
\be
    B_x  =  \frac{\mu_0 j_y h}{2}  =  \frac{\mu_0 I}{2 w}, 
\ee
where $\mu_0$ is the vacuum permeability, \hl{$h$} is the thickness of the sample, and $w$ is the width of the sample. Given the sample dimensions listed in Supp. Sec. A, we expect a field-per-current coefficient of \SI{3.7}{\gauss / \A},  in agreement with the BEC-height measurements. This shows that the local $j_{\textrm{bulk}}$ at the location we perform our magnetometry does not deviate substantially from that calculated assuming a spatially uniform distribution.

\subsection{Extracting current density from magnetic field}
\label{ssec:deconv}

We describe the method used to extract current density $j_y(x,y)$ from the magnetic field $B_x(x,y)$ measured using the SQCRAMscope. This method expands upon that detailed in Ref.~\cite{Yang:2017br}. Assuming an infinite sheet of electric current that is uniform along its thickness $h$, the Green's function for the Biot-Savart kernel $G(k_x, k_y)$ used to compute the field a distance $r$ from the surface of the sheet is given by
\begin{align}
\label{equ:bskernel}
    G(k_x, k_y) = \mu_0 \sinh{(h \bar{k}/2)} \exp{\left[-\bar{k} (d+h/2)\right]} / \bar{k} ,
\end{align}
where $\bar{k} \equiv \sqrt{k_x^2+k_y^2}$ is the spatial wavenumber. The $y$-component of the current density $j_y$ is computed by deconvolution with the Biot-Savart kernel, or equivalently by division in Fourier space:
\begin{align}
    j_y(k_x, k_y) = B_x(k_x, k_y)/G(k_x, k_y).
\label{eq:jydeconv}
\end{align}

\begin{figure}
    \centering
    \includegraphics[width=0.8\linewidth]{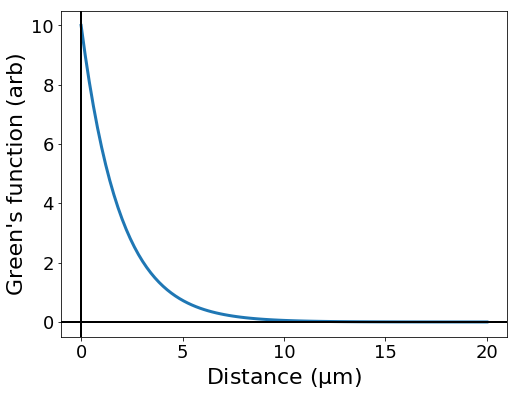}
    \caption{Green's function versus depth below sample surface for a 12-$\mathrm{\mu m}$ spatial wavelength roughly corresponding to the observed domain length scale.}
    \label{supfig:bulk_sensitivity}
\end{figure}

The Green's function for an infinitesimally thin sheet of current a depth $d$ within the sample decays exponentially with length scale $1/\bar{k}$. Thus, spatial frequencies corresponding to \SI{12}{\um}, for example, which close to the width of a typical domain in our samples, have a decay length of $1/\bar{k} = \SI{2}{\um}$ within the bulk. Supplementary~Fig.~\ref{supfig:bulk_sensitivity} plots the Green's function versus depth below the sample surface for this spatial frequency. It shows that magnetometry is primarily sensitive to the top few microns of sample current when detecting signals with spatial extent matched with the typical domain size. In contrast to the \SI[number-unit-product=-]{30}{\nm} penetration-depth scale of the optical measurements, this corresponds to bulk length scales. 

The convolution method described above is mathematically exact, but also extremely sensitive to high-frequency noise due to the exponential term in Eq.~\eqref{equ:bskernel} unless $\Gamma~\equiv~(d \pm h/2) \max{(\bar{k})} \ll 1$. In the present work $\Gamma \approx 300 $. Thus, we must suppress high frequencies with an appropriately chosen window function. We use a Hanning window,
\be
    H(\bar{k}) = 
    \begin{cases} 
      \cos{}^2 \frac{\bar{k}}{2 \lambda} & \bar{k} \leq 2 \pi / \lambda \\
      0 & \bar{k} > 2 \pi /\lambda
   \end{cases}
   ,
\label{eq:windowfn}
\ee
with a cut-off that removes all frequencies greater than $2 \pi / \lambda$. 
The value of $\lambda$ should be chosen large enough to not filter out critical frequency components of the signal, but not so large as to allow excessive amounts of noise to corrupt the signal. This number will set the effective spatial resolution of the SQCRAMscope for imaging current density, down to a limit no smaller than the spatial resolution for magnetic field imaging (presently \SI{2.2}{\um} with the lens system being used)~\cite{Yang:2017br}. For the 2D current density plots in Figs.~2 (g) and (h), the large thickness of the samples requires a relatively large value of $\lambda$ to adequately suppress high-frequency noise. We choose $\lambda=8~\mu$m, resulting a FWHM point-spread resolution of \SI{8}{\um} when imaging current density in these samples. The 8-$\mu$m cutoff is chosen to minimize the amplified noise without significantly reducing the size of the measured signal  in any but the smallest of domains;  \SI{8}{\um} is close to the width of the narrowest domains imaged in Figs.~2.

\subsection{Model relating  current density to resistivity anisotropy}
\label{ssec:ra_model}

\begin{figure}
    \centering
    \includegraphics[width=0.99\linewidth]{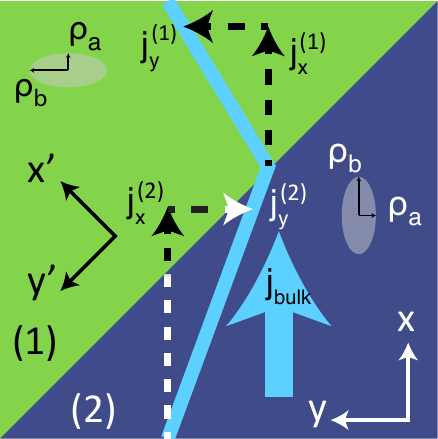}
    \caption{Domain structure for theoretical calculation of anisotropy.}
    \label{fig:anis_theory}
\end{figure}

This section describes the derivation of an expression for resistivity anisotropy. An analytic model is used to relate resistivity anisotropy to current density. We then use this model in  Supp.~Sec.~\ref{ssec:ra_from_current} to calculate resistivity anisotropy from measured current density.  

Consider an infinite conductor in a 2D plane representing a crystal in the orthorhombic phase. A single, infinitely long domain wall extends through the crystal at \ang{45} to the $x$ and $y$ axes, as shown in Supp.~Fig.~\ref{fig:anis_theory}. In the upper domain, denoted (1), the $a$ and $b$ crystal axes are parallel to the $x$ and $y$ coordinate axes, respectively, while in the lower domain, denoted (2), the situation is reversed: $a$ is parallel to $y$ and $b$ is parallel to $x$. In each domain, the resistivity takes a value $\rhoa$ along the crystal $a$ axis and $\rhob$ along the $b$ axis.

In this model, current is driven along the $x$ direction, and the current flow is deflected at the domain boundary by the anisotropic resistance to give a finite current density in the $y$ direction. Let $\jxone$, $\jxtwo$, $\jyone$, and $\jytwo$ refer to current densities in the $x$ and $y$ directions in either the first or second domain region, as denoted by the superscript. We can experimentally measure the difference in current along the $y$ direction in the two domains
\begin{align}
\jmeas \equiv \jytwo-\jyone.
\label{eq:jmeas}
\end{align}
We treat all other current densities, as well as the resistivities, as unknown, and gather a set of equations that will let us solve for the ratio of resistivities, $\drho$.

Two relations can be obtained by conservation of charge and Faraday's law. These give, respectively, $\nabla \cdot \vec{j} = 0$ and $\nabla \times \vec{E} = 0$, where $\vec{E}$ denotes the electric field. Taking a divergence and line integral, respectively, across a long, narrow box straddling the domain boundary, we can convert these differential equations to simple forms. Defining the coordinates $x'$ ($y'$) to be perpendicular (parallel) to the domain wall, we write the result of these integrals as:
\bea
\jxpone &=& \jxptwo\\
E^{(1)}_{y'} &=& E^{(2)}_{y'}.
\eea
Converting back to the $x-y$ coordinate system and inserting the constitutive
equation $E_{i} = \rho_{i j} j_j$, we arrive at the following equations:
\bea
\jxone + \jyone &=& \jxtwo + \jytwo
\label{eq:charge}\\
-\rhoa \jxone + \rhob \jyone &=& -\rhob \jxtwo + \rhoa \jytwo.
\label{eq:maxwell}
\eea

We now obtain a final set of two equations by inserting assumptions about the net flow of current. To obtain a more general result, we allow for the two domains to be of unequal size, letting a fraction $f_1$ of the sample be domain 1 and a fraction $f_2$ be domain 2, where $f_1 + f_2 = 1$. We assume there is a net current density $\jbulk$ in the $x$ direction, and no net current in the $y$ direction, representing our current supply driving electronic transport through the crystal. By averaging the $x$ current over a line parallel to the $y$ axis, while averaging the $y$ current over a plane parallel to the $x$ axis, we obtain:
\bea
f_1 \jxone + f_2 \jxtwo &=& \jbulk
\label{eq:jbulk}\\
f_1 \jyone + f_2 \jytwo &=& 0.
\label{eq:jperp}
\eea

We can now solve this set of equations for $\drho$ in terms of $\jmeas$, $\jbulk$, $f_1$, and $f_2$. First, we take Eq.~\eqref{eq:jmeas} and Eq.~\eqref{eq:jperp}, which combine to give:
\bea
\jyone &=& -f_2 \jmeas
\label{eq:jxone}\\
\jytwo &=& f_1 \jmeas.
\label{eq:jxtwo}
\eea
Substituting these into Eq.~\eqref{eq:charge} and Eq.~\eqref{eq:jbulk}, we solve for $\jxone$ and $\jxtwo$:
\bea
\jxone &=& \jbulk + f_2 \jmeas\\
\jxtwo &=& \jbulk - f_1 \jmeas.
\eea
Finally, we substitute the above into Eq.~\eqref{eq:maxwell}:
\begin{align}
-\rhoa \left(\jbulk + f_2 \jmeas\right) - \rhob f_2 \jmeas = \\
-\rhob \left(\jbulk - f_1 \jmeas\right) + \rhoa f_1 \jmeas.
\end{align}
Simplifying, we obtain:
\begin{align}
\rhoa/\rhob = \frac{\jbulk - \jmeas}{\jbulk + \jmeas}.
\end{align}
We can then rewrite this as
\begin{align}
\drho = \frac{2 \jmeas}{\jbulk + \jmeas}.
\label{eq:anis}
\end{align}
This equation provides the resistivity anisotropy as a function of only the known bulk current density $\jbulk$ and the measured current density $\jmeas$. Note that this equation is also independent of the relative size of the domains, as the geometric factors $f_1$ and $f_2$ do not appear in the result.

To verify the validity of this equation, we performed finite element simulations of electric current flowing through adjacent domains. The domains have alternating anisotropic resistance and varying widths, similar to the domain patterns we see in the measured \BFCA~crystals. Equation~\eqref{eq:anis} correctly determined the anisotropy in these models to within numerical error of a few percent, and did so consistently for a variety of domain widths and anisotropy magnitudes.

\subsection{Computation of resistivity anisotropy from magnetic field}
\label{ssec:ra_from_current}

We now detail the use of the model described above in Supp.~Sec.~\ref{ssec:ra_model} to compute the temperature dependence of the resistivity anisotropy from the measured magnetic field $B_x$. To compute the resistivity anisotropy using Eq.~\eqref{eq:anis}, one must first calculate the current density $j_y$. While the deconvolution method detailed in Supp.~Sec.~\ref{ssec:deconv} calculates $j_y$ from $B_x$ using minimal assumptions about the spatial structure of $j_y$, it is susceptible to making a biased estimate of the $j_y$ modulation amplitude due to the need to choose a low-pass filter cut-off frequency. In Fig.~2 of the main text we establish, using the deconvolution method, that the domain structure exhibited in $j_y$ is in good correspondence with that in the birefringence. Thus, the birefringence signal provides prior knowledge of $j_y$ that may be used to make an estimate of $j_y$ from $B_x$ that is less susceptible to bias.  

Rather than computing $j_y$ by direct deconvolution of $B_x$, we use an iterative method. Using birefringence images, we construct a parametric model of $j_y$ which is then convolved with the Biot-Savart kernel Eq.~\eqref{equ:bskernel} to yield a trial magnetic field $B'_x$. We then vary the model parameters by gradient descent so as to minimize the residual squared error (RSE) between $B'_x$ and the measured magnetic field $B_x$. The amplification of high-frequency noise discussed in Supp.~Sec.~\ref{ssec:deconv} is avoided because this method does not directly deconvolve magnetic field data. The peak-to-peak amplitude of $j_y$, a model parameter, can then be used to compute anisotropy as described in Supp.~Sec.~\ref{ssec:ra_model}.

\begin{figure}
    \centering
    \includegraphics[width=0.99\linewidth]{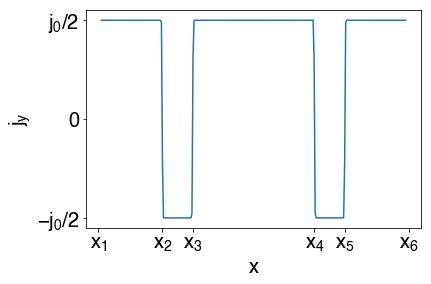}
    \caption{Example of current density model used to fit magnetic field data.}
    \label{supfig:j_model}
\end{figure}

As shown in Supp.~Fig.~\ref{supfig:j_model}, we model $j_y$ as being of fixed magnitude but rapidly reversing polarity upon crossing a domain wall. We parameterize this model according to
\be
    j_y(x, j_0, \epsilon, x_1, \ldots, x_N) = \sum_{i=1}^{N} (-1)^i j_0 \left[\text{erf} ([x-x_i]/\epsilon) + 1 \right]/2 ,
\ee
where $j_0$ is the amplitude of the current; the $x_i$ define the positions of the domain walls, and $\epsilon \to 0$  such that the error function approximates a step function. 

The trial magnetic field $B'_x$ is computed as the convolution of $j_y(x,j_0, x_1, \ldots, x_N)$ with the Biot-Savart Green's function $G$:
\bea
       B'_x(x, j_0, x_1, \ldots, x_N, B_0, B_1) =\,\, & &G*j_y(x, \epsilon, x_1, \ldots, x_N) \nonumber\\
       &&+B_0+B_1 x + B_2 x^2,
\eea
where $B_0$, $B_1$, and $B_2$ account for fields produced by large-length-scale current modulations near the region of interest. Choosing a temperature where the magnetic signal is near its strongest, we determine the position of the domain walls by minimizing the RSE between $B'_x$ and $B_x$ by varying $j_0$, $x_i$, $B_0$, and $B_1$. We then constrain the $x_i$ such that the spacing between domain walls is fixed but they may undergo a rigid translation and the position of the domain walls is determined by a single fit parameter $x_0$. To determine the temperature dependence of $j_y$ we use the trial magnetic field
\bea
    B'_x(x, j_0, x_0, B_0, B_1) =\,\, &&  G*j_y(x, \mu, x_1+x_0, \ldots, x_N+x_0)\nonumber\\
    &&+ B_0 + B_1 x + B_2 x^2,
    \label{eq:Bxfit}
\eea
to find the $j_0$, $x_0$, $B_0$, $B_1$, and $B_2$ that minimize the RSE in magnetic field. We can then use Eq.~\eqref{eq:anis} to compute the resistivity anisotropy at each temperature by setting $\jmeas=j_0$.

\begin{figure*}[t!]
    \centering
    \includegraphics[width=0.99\linewidth]{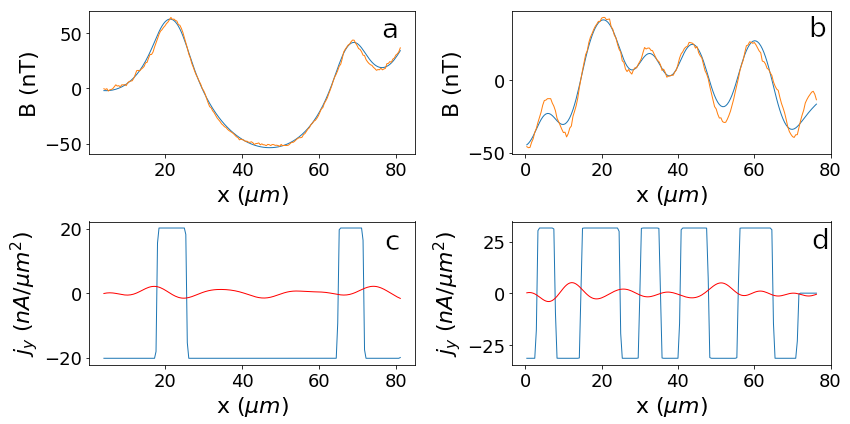}
    \caption{Error estimation. Panels (a) and (b) show the typical measured magnetic field (orange) and the magnetic field that results from the optimized current density model (blue) for the parent and doped samples,respectively. Panels (c) and (d) show the optimized current density (blue) and current density error (red) produced by deconvolving the difference in measured and calculated magnetic field in panels (a) and (b). }
    \label{supfig:errors}
\end{figure*}

We estimate the uncertainty in current density by comparing the measured magnetic field to that calculated using  the current density model and its optimal set of parameters. Supplementary~Figs.~\ref{supfig:errors} (a) and (b) show typical measured (orange) and model (blue) magnetic field for the parent and 2.5\%-doped samples, respectively. We take the difference of these two curves and deconvolve the result with the Biot-Savart kernel, following the procedure in Supp.~Sec.~\ref{ssec:deconv} with a cutoff frequency of 8~$\mu$m. The resulting current density represents an estimate of the difference between our model current density and the current density flowing through the sample. We overlay in Supp.~Fig.~\ref{supfig:errors} (c) and (d) for the parent and 2.5\%-doped samples, respectively, the model (blue) and error (red) current densities for the same data presented in (a) and (b). 

We define the uncertainty in $\jmeas$ to be the spatial standard deviation in the computed error current density added in quadrature with error resulting from a 10\% variation in sample thickness. The resistivity anisotropy error bars in Figs.~3 (e) and (f) result from propagating these uncertainties through Eq.~\eqref{eq:anis}. These error bars represent uncertainty due to both random sources (e.g., noise in the measurement) and systematic sources, such as a specification of the current density model. 

\subsection{Definition of domain-averaged amplitudes}
\label{ssec:domain_amp}

We now provide the definitions for the domain-averaged amplitudes $A_{B}(T)$ and $A_{\theta}(T)$ used in the main text. We define the domain-averaged amplitude for magnetometry and birefringence modulations to be
\begin{equation}
A_{B}(T) = \argmin_{\alpha} \left \{ \int \textrm{d}x \left [\bar{B}_x(x,T) - \alpha \bar{B}_x(x,T_{\textrm{ref}})\right ]^2 \right \} 
\end{equation}
and
\begin{equation}
A_{\theta}(T) = \argmin_{\beta} \left \{ \int \textrm{d}x \left [\bar{\theta}_x(x,T) - \beta \bar{\theta}_x(x,T_{\textrm{ref}})\right ]^2 \right \},
\end{equation}
respectively. $T_{\textrm{ref}}$ is a reference temperature chosen to be that where the amplitude of spatial modulations associated with domains is largest.
We expect the domain amplitude to decrease from a peak value near 1 down to 0 as temperature rises through $T_{\text{nem}}$. The domain-averaged amplitude will predominately reflect the size of features that are large in amplitude or extent. Smaller or unresolved features will therefore have a minimal effect on the amplitude, and the presence of, e.g., narrow domains that are visible in birefringence but not visible in magnetometry, will not negatively impact the efficacy of this technique.

\subsection{Simulation of magnetic field for Figure~1d}
\label{ssec:sim_field}

We simulated the magnetic field we expect to measure in a two-step process. First, the  current density for a given configuration of nematic domains was computed using finite-element analysis.  This current density was then used to numerically compute the magnetic field by convolution with the Biot-Savart kernel; see~Supp.~Sec.~\ref{ssec:deconv}.

\subsection{Birefringence measurements}
\label{ssec:polarimeter}
\begin{figure}
    \centering
    \includegraphics[width=0.99\linewidth]{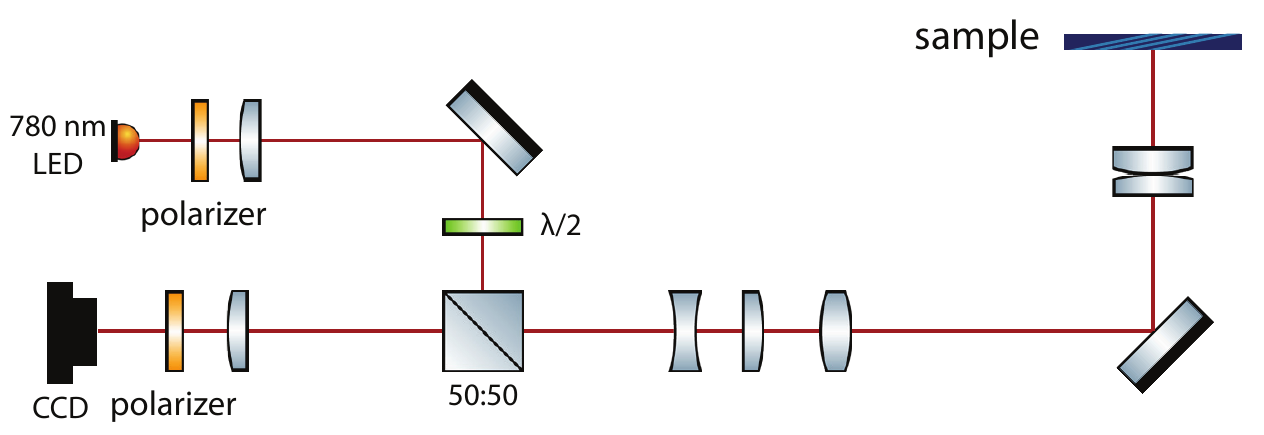}
    \caption{Polarimeter schematic for optical birefringence measurements.}
    \label{supfig:pol_optics}
\end{figure}

We  augmented our SQCRAMscope magnetometer with an optical birefringence microscope similar to the setup in~\cite{Tanatar:2009ks}; see Supp.~Fig.~\ref{supfig:pol_optics}. The sample, mounted on a silicon substrate in a UHV chamber, is illuminated by a \SI[number-unit-product=-]{780}{\nm} LED  with polarization set by a linear polarizer  and a $\lambda/2$ waveplate. Light reflected from the sample passes through another linear polarizer  and is imaged onto a CCD camera. Silver-coated mirrors and a 50:50 plate beamsplitter are carefully chosen to minimize distortion of the polarization, giving rise to an extinction ratio in excess of 1:1000.  This provides an  angular resolution better than \ang{0.1}. The imaging optics are specifically designed to be installed in the SQCRAMscope with in-vacuum lenses to provide better numerical aperture. The microscope was tested with a 1951 USAF target and found to have a spatial resolution of ${\sim} 3$~$\mu$m, estimated using the Rayleigh criterion.

The microscope determines the polarization rotation of light reflected from the sample using the two nearly crossed polarizers. The intensity of the light on the CCD camera is therefore indicative of the rotation angle $\Delta \theta = \theta_{\textrm{out}} - \theta_{\textrm{in}}$, where $\theta_{\textrm{in}}$ is the linear polarization angle of the incident light and $\theta_{\textrm{out}}$ is that of the reflected light. For samples discussed in the main text, the largest signal is found when the incident light is linearly polarized along the orthorhombic (110)$_{o}$ direction. This is consistent with the breaking of $C_4$ symmetry resulting in a   reflectance  difference between light polarized along (100)$_{o}$ versus (010)$_{o}$. Therefore, we fix the incident polarization to be at \ang{45} with respect to the orthorhombic axes, and define it to be the angular origin throughout the paper, i.e., $\theta_\textrm{in}=0$ and $\theta_\textrm{out}=\Delta \theta$. 

Because only a small optical birefringence is exhibited by \BFCA, even in the orthorhombic state, care must be taken to observe a signal. Depending on whether high accuracy or precision is required, we choose to operate the optical birefringence microscope in one of two imaging modes. 

The first mode, which we call the relative mode, enables precise measurements of the relative polarization rotation angle between points on the sample $\Delta \theta(\vec{r}_1) - \Delta \theta(\vec{r}_2)$ at the expense of an overall angular offset. For a given location on the sample, the intensity recorded on the camera is
\begin{equation}
    I(x, y; \alpha) \propto \sin^2[\alpha - \Delta \theta(x, y)],
\end{equation}
where we denote the angle of the second polarizer (the analyzer) measured from the maximum extinction position in the absence of birefringence $\alpha$. 
By recording images at a series of $\alpha$ values, $\Delta \theta$ is extracted through a least-square fit to $I(\alpha)$. However, we note that the accuracy of $\Delta \theta$ is limited by that of $\alpha$, which is set by the precision of optical components and is on the order of \ang{0.2}.  Consequently, small $\Delta \theta$ cannot be directly compared against zero to infer the sign of polarization rotation. This mode of operation is suitable for measuring the contrast between twin domains, and therefore was employed for the data shown in Figs.~2--4.

The second mode, which we call the absolute mode, obviates this problem by fixing the angle of the analyzer $\alpha$ during sample cool-down and warm-up. This provides an improved accuracy in angle measurements but suffers from reduced precision. In this operation mode, a differential image is constructed as
\begin{eqnarray}
    \Delta I(x, y) & = & I(x, y; \alpha_+) - I(x, y; \alpha_-) \nonumber\\
                   & \propto & \sin^2(\alpha_+ - \Delta \theta) - \sin^2(\alpha_- - \Delta \theta)\nonumber \\
                   & \approx & (\alpha_+^2 - \alpha_-^2) - 2 (\alpha_+ - \alpha_-) \Delta \theta(x, y),
\end{eqnarray}
where $\alpha_-$ and $\alpha_+$ are the angles of the analyzer during cool-down and warm-up, respectively. Here, the previous uncertainty in $\alpha$ is replaced by the unknown but fixed proportionality factor  $\Delta \theta(x, y)$ across datasets. We are therefore able to compare the rotation angle directly to zero and thus identify domains of opposite sign in the nematic order parameter. In addition, with a reference dataset taken in the relative mode, we are able to calibrate-out the unknown proportionality factor. This results in the dataset shown in Supp.~Fig.~\ref{fig5}, where the birefringence of both parent and 2.5\%-doped samples is measured across a large temperature span, all while retaining angular resolution in absolute units.

\subsection{Registration of optical birefringence and magnetometry images}%

The optical birefringence and the magnetometry scans are performed using different optical axes and imaging systems. To remove this spatial offset of the birefringence images to the magnetometry maps, we introduce a linearly polarized \SI[number-unit-product=-]{780}{\nm} laser beam resonant with the $^{87}$Rb D2 transition along the same path as the polarimetry light source (see white beam in Fig.~1) and perpendicular to the sample surface. Absorption images of the BEC are then collected on the birefringence imaging camera. The same BEC is also imaged through the SQCRAMscope imaging axis (see red beam in Fig.~1). Together with the known magnification of the two imaging systems, this allows us to construct the coordinate transformation that brings the magnetometry data into the same coordinate system as the birefringence images.    \hl{In particular, we take images of the samples with the birefringence imaging camera immediately after taking each BEC absorption image on the SQCRAMscope imaging axis. An up-sampled DFT cross-correlation-based image registration algorithm~\cite{Padfield:iz,GuizarSicairos:2008vj} is used to register the optical sample images so as to reconstruct the magnetometry scan regions on the sample plane. This provides us with the ability to make a direct comparison between the two modes of operation, as shown in Figs.~2--4 of the main text. We further apply the cross-correlation method between datasets taken at different temperatures to align  features seen in magnetometry of Figs.~2 and 4.}

\subsection{Trap parameters}

\hl{The atomic cloud utilized by the SQCRAMscope is confined by a magnetic Ioffe-Pritchard trap as described in previous work~\cite{Yang:2017br}. In the present work, the longitudinal and transverse magnetic traps have typical frequency \SI{12.2 \pm 0.2}{\Hz} and \SI{1.41 \pm 0.05}{\kilo\Hz}, respectively. In addition, a uniform \SI{0.3}{\milli\tesla} bias field provides an atomic quantization axis along $\hat{x}$. The trap is positioned $2.3 \pm 0.4$-$\mu$m below the surface of the sample. The atoms could be positioned as close as 800~nm~\cite{Yang:2017br}, limited by trap width, but we choose a large distance in this work for ease of use.  }

\hl{The magnetic field experienced by the sample due to the magnetic trap is close to the $0.3$-mT trap bias field near the atomic cloud, and increases to a maximum value of less than 10~mT at the edge of the sample farthest from the atoms. Magnetic fields of these small magnitudes are expected to have a negligible effect on the sample's resistivity and transition temperature~\cite{Chu:2010hg}.}
%
%
%

\end{document}